\documentclass[amsmath,amssymb,pra,aps,mathbbm,superscriptaddress]{revtex4}
\usepackage{amsmath,amsfonts,amssymb,amsthm,graphics,graphicx,epsfig,bbm}
\usepackage[colorlinks=true,citecolor=blue,linkcolor=blue,urlcolor=blue]{hyperref}
\usepackage[usenames]{color}
\usepackage{graphicx}
\usepackage{subfigure}
\usepackage{epsfig}
\usepackage{bbold}
\usepackage{dcolumn}
\usepackage{bbm}
\usepackage{color}
\usepackage{verbatim}
\usepackage{epstopdf}
\usepackage{amssymb} 
\usepackage{amstext}
\usepackage{latexsym}
\usepackage{hyperref}
\usepackage{amsfonts}
\usepackage{psfrag}
\usepackage{xcolor}
\usepackage{times}
\usepackage{mathtools}

\begin{document}
\title{Quantum illumination reveals phase-shift inducing cloaking}
\author{U. Las Heras}
\affiliation{Department of Physical Chemistry, University of the Basque Country UPV/EHU, Apartado 644, E-48080 Bilbao, Spain}
\author{R. Di Candia}
\affiliation{Department of Physical Chemistry, University of the Basque Country UPV/EHU, Apartado 644, E-48080 Bilbao, Spain}
\affiliation{Dahlem Center for Complex Quantum Systems, Freie Universit\"at Berlin, 14195 Berlin, Germany}
\author{K. G. Fedorov}
\affiliation{Walther-Meissner-Institut, Bayerische Akademie der Wissenschaften, D-85748 Garching, Germany}
\affiliation{Physik-Department, Technische Universit\"at M\"unchen, D-85748 Garching, Germany}
\author{F. Deppe}
\affiliation{Walther-Meissner-Institut, Bayerische Akademie der Wissenschaften, D-85748 Garching, Germany}
\affiliation{Physik-Department, Technische Universit\"at M\"unchen, D-85748 Garching, Germany}
\affiliation{Nanosystems Initiative Munich (NIM), Schellingstraße 4, 80799 M\"unchen, Germany}
\author{M. Sanz}
\email{mikel.sanz@ehu.eus}
\affiliation{Department of Physical Chemistry, University of the Basque Country UPV/EHU, Apartado 644, E-48080 Bilbao, Spain}
\author{E. Solano}
\affiliation{Department of Physical Chemistry, University of the Basque Country UPV/EHU, Apartado 644, E-48080 Bilbao, Spain}
\affiliation{IKERBASQUE, Basque Foundation for Science, Maria Diaz de Haro 3, 48011 Bilbao, Spain}

\begin{abstract} 

In quantum illumination entangled light is employed to enhance the detection accuracy of an object when compared with the best classical protocol. On the other hand, cloaking is a stealth technology based on covering a target with a material deflecting the light around the object to avoid its detection. Here, we propose a quantum illumination protocol especially adapted to quantum microwave technology. This protocol seizes the phase-shift induced by some cloaking techniques, such as scattering reduction, allowing for a $3$ dB improvement in the detection of a cloaked target. The method can also be employed for the detection of a phase-shift in bright environments in different frequency regimes. Finally, we study the minimal efficiency required by the photocounter for which the quantum illumination protocol still shows a gain with respect to the classical protocol.
\end{abstract}

\date{\today}

\maketitle

The idea of covering an object with a cloak to render it invisible is an old dream in literature and science fiction. However, it was not until the 1960's that specific theoretical proposals emerged~\cite{Dolin, Kerker}. The key paradigms of cloaking are based on the reduction or cancellation of the scattering cross section of an object by modifying the surrounding electromagnetic field, or steering of incident illumination around an object and reforming it on the other side. Although physical implementations of cloaking were impossible for many years, the last decade had seen a dramatic improvement in cloaking technology~\cite{Leonhardt, Pendry, Schurig} due to advances in metamaterials~\cite{Liu,Lapine}. Making use of these metamaterials, light reflected off of a background illuminates the cloaked object, which diverts and reconstructs afterwards the wavefront with the same trajectory. As a consequence, an observer only sees the background without noticing the object. Currently, several methods for cloaking, such as coordinate transformation~\cite{Pendry, Schurig}, transmission line~\cite{Alitalo2, Alitalo3} and metal-plate~\cite{Tretyakov} cloaks, plasmonic media~\cite{Alu}, have been proposed and experimentally compared~\cite{Alitalo}, as well as carpet~\cite{Li}, exterior~\cite{Lai} and space-time~\cite{McCall} cloaks, besides illusion generators~\cite{Jiang} and acoustic cloaks~\cite{Nelson}.  Benefits and handicaps of each method have been analyzed in detail, evaluating reflectivity coefficients and phase shifts generated in the wavefront due to the cloak. Furthermore, in the last few years, it has been shown that it is possible to cloak targets in a different range of frequencies~\cite{Shchelokova}, including microwave~\cite{Schurig,Liu15}, terahertz~\cite{Zhou} and optical regimes~\cite{Ergin}.

Quantum illumination~\cite{Lloyd} utilizes quantum properties such as superposition and entanglement in order to detect the presence of a low reflectivity object in a noisy environment with a higher accuracy than any protocol employing classical light. In general, the idea consists in preparing a pair of entangled optical beams and irradiating the target with one of them, while preserving the other one in the lab. In comparison to classical light, the existence of quantum correlations allows us to declare the presence of the object with either a higher accuracy or less resources, achieving more than 3dB using collective or adaptive measurements, which are extremely challenging with current technology~\cite{Zhuang}. The relevant figure of merit in this context is the signal-to-noise ratio (SNR)  of the respective protocols. For optical frequencies, methods making use of multi-photon beams~\cite{Saphiro}, gaussian states~\cite{Tan}, optical receivers~\cite{Guha}, and photon subtraction operators~\cite{Zhang} have been proposed, as well as experimental realization of quantum illumination protocols measuring photon-number correlations~\cite{Lopaeva} and demonstrating quantum-illumination-based secure communication~\cite{Zhang13} and sensing enhancement~\cite{Zhang15}. Research in quantum illumination has essentially focused on the detection of low-reflectivity objects, leaving apart phase shift detection. Nonetheless, the use of quantum properties to detect phase shifts has been addressed in other contexts, particularly in quantum estimation methods~\cite{Sanders97} and in quantum state discrimination techniques for quantum communication~\cite{Muller15}, but generally in non-diffusive environments. Recently, quantum illumination protocols have been adapted to the microwave spectrum~\cite{Barzanjeh}, where classical protocols are known to be adequate for the detection of macroscopic objects. Indeed, radar systems make use of RF electromagnetic waves in order to locate targets due to the object size. Additionally, the microwave regime is particularly relevant, since the atmosphere is mostly transparent at frequencies around 20~GHz~\cite{Sharkov}. In this range, the atmospheric interaction with beams can be neglected, while infrared beams can easily be absorbed or scattered. There are numerous motivations for extending ideas from quantum illumination into a fully fledged quantum radar~\cite{Lanzagorta}, including space exploration, airspace control, and maritime radar applications, among others.

In this Article, we demonstrate how to exploit entanglement in order to detect a cloaked object with higher accuracy than any classical protocol. Cloaking methods are imperfect, since while trying to minimize photon losses, they introduce a phase shift after the deflection of the wavefront~\cite{Alitalo}. We exploit this effect to detect the presence of the object. In particular, we consider the realistic scenario of illuminating the target with a wavefront previously reflected by a noisy reference background. This interaction is simulated by a high-reflectivity mirror merging the signal with a bright thermal state modeling the noisy environment. We quantify the performance of the proposed quantum protocol in comparison to the best classical one, where the employed light can be described in terms of coherent states, by explicitly calculating the SNR for both of them and find a significant improvement of up to 3~dB. Furthermore, we propose a specific implementation of our protocol at microwave frequencies making use of a Josephson mixer~\cite{Flurin} and photocounters. Finally, we calculate the minimal photocounter efficiency required for the quantum illumination to be still advantageous.

Given that metamaterial cloaking achieves extremely low photon losses in the microwave regime~\cite{Schurig}, searching for new measurable properties is essential. Particularly, small phase disturbances have been observed in coordinate transformation, transmission line, and metal-plate cloaks~\cite{Alitalo}. In this sense, it is necessary to discover the measurement maximizing the distinguishability between the cases in which the cloaked object is present or absent. Actually, the optimal protocol for measuring a phase shift is based on interferometry, which is not implementable in our model, due to the interaction with the thermal environment.

\begin{figure*}[t] 
\begin{center}
\includegraphics[width=0.9\textwidth]{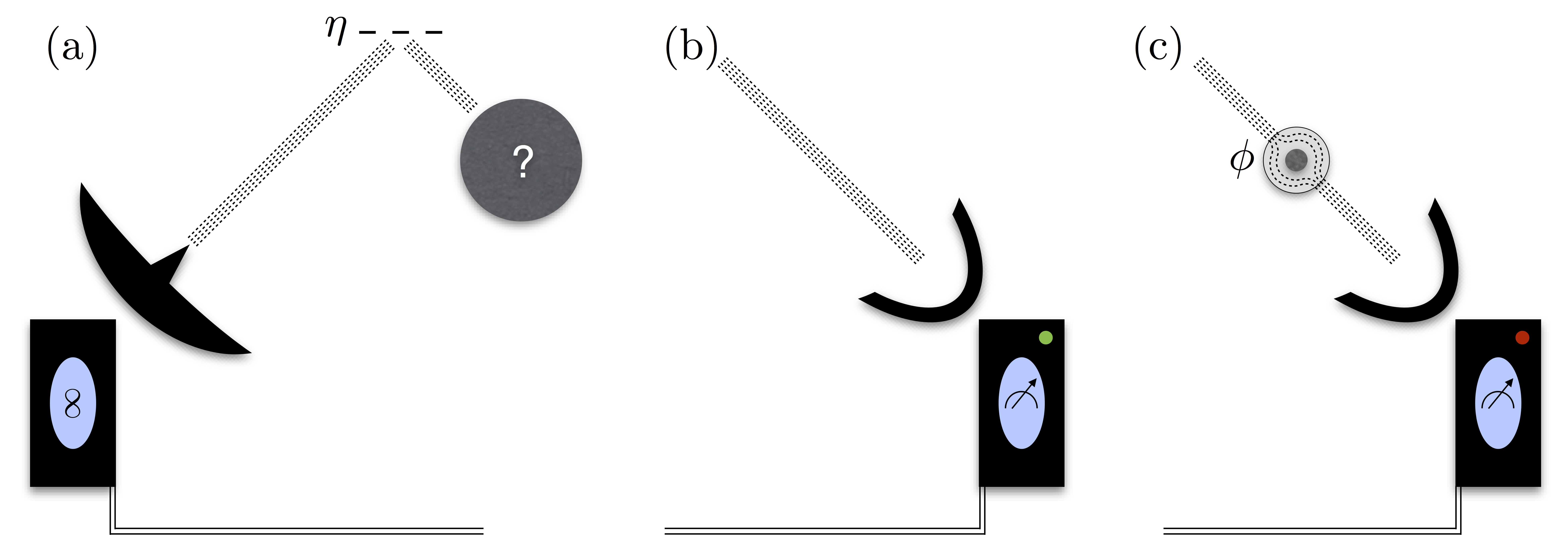}
\end{center}
\caption{Pictorial diagram of the quantum protocol for the detection of cloaked targets. (a) A two-mode squeezed state is generated in the lab. The idler beam stays in a controlled transmission line while the signal is emitted toward a highly reflective background, $\eta\approx1$, where is mixed with thermal noise. When the signal comes back, it either can be directly received by the measurement device (b) or it can pass before through a metamaterial-cloak which covers a target we want to detect (c) In the latter case, a phase shift is introduced due to the imperfection in current cloaking elements. In any case, a bipartite operation is performed with a Josephson mixer onto the signal and idler beams. Finally the number of photons in the idler beam is measured, which gives the information about the presence of the cloaked object with a gain up to 3~dB over the classical protocol.}
\label{Scheme}
\end{figure*}

Let us describe the elements of the protocols. We make use of a coherent state $|\alpha\rangle$ and a Gaussian two-mode squeezed state $|\Psi_{12}\rangle$ for the classical and the quantum protocols, respectively. For the latter, while the idler remains in the lab, the signal beam is partially reflected off from the reference high-reflectivity background, which can be written in terms of the photon field as follows
\begin{eqnarray}\label{Eq1}
a'=\sqrt{\eta}a+\sqrt{1-\eta}a_{h},
\end{eqnarray}
where $a$ $(a')$ is the bosonic annihilation operator of the incoming (outcoming) field in the signal, $a_h$ the annihilation operator of the environmental field, and $\eta$ is the reflectivity coefficient of the reference background. Hence, by this mechanism, the signal beam is mixed with the noisy environment, which is defined as a thermal state $\rho_{th}=(1-\lambda_{th})\sum_{n=0}^\infty\lambda_{th}^n|n\rangle\langle n|$, with $n_{th}=\frac{\lambda_{th}}{1-\lambda_{th}}\gg1$. In the limit $\eta=1$ it would work as a perfect mirror. After this step, the signal beam is deflected by the cloak of the target. For the sake of simplicity, we consider that the phase change $\phi$ introduced by the imperfect cloak yields a transformation in the field given by
\begin{eqnarray}\label{Eq2}
a''=a'e^{-i\phi},
\end{eqnarray}
where $a''$ is the field after it has been deflected from the cloak. In case of a coherent state $|\alpha\rangle$, the outcoming state would be $|\alpha e^{-i\phi}\rangle$. Finally, the signal beam is received in the lab and a joint measurement is performed to detect a phase shift $\phi\neq0$. Accordingly, the SNR is calculated, since this gives the improvement in the Chernoff bound determining the presence of the object~\cite{Zhang,Sanz}.

\paragraph{Classical protocol.---}In the following, we discuss the optimal classical protocol to detect a small phase shifts. A coherent state is prepared in the lab~\cite{Fedorov}
\begin{eqnarray}
|\alpha\rangle=e^{-|\alpha|^2/2}\sum_{n=0}^\infty\frac{\alpha^n}{\sqrt{n!}}|n\rangle.
\end{eqnarray}
Once the state is transformed according to Eqs.~\eqref{Eq1} and~\eqref{Eq2}, the quadrature in the angle of the coherent state is measured, which expectably gives the optimal SNR~\cite{Tan, Guha, Sanz} in the limit $\eta\sim1$. Without loss of generality, we consider $\alpha\in R$ and then the quadrature measurement is just $\langle x''\rangle$, a prototypical measurement in microwaves~\cite{Fedorov,Fedorov2,DiCandia}. Therefore, the SNR in the classical protocol is given by the expression
\begin{eqnarray}\label{Eq4}
\left(\frac{S^2}{\sigma^2}\right)_C=\frac{\langle \bar x'' \rangle^2}{\langle \bar x''^2 \rangle-\langle \bar x'' \rangle^2}=\frac{2\eta N(1-\cos(\phi))^2}{\eta/2+(1-\eta)(n_h+1/2)}.
\end{eqnarray}
Here, $\bar x=\langle x\rangle|_{\phi=0} - \langle x\rangle$, so that the SNR is zero when the dephased angle $\phi=0$, $\eta$ is the background reflectivity, $n_h$ is the number of thermal photons, and $N=|\alpha|^2\ll1$ is the number of photons in the initial coherent signal. See supplementary information for detailed calculations. From this expression, one can see that SNR grows with the number of photons in the signal beam, and decreases with the environmental noise. Note that the same result is achieved by considering the presence of the object before the reflection of the signal from the blackground.

\paragraph{Quantum protocol.---} Let us analyze the improvement generated by a protocol using an entangled state, as depicted in Fig.~\ref{Scheme}. For this, we generate a two-mode squeezed state~\cite{Menzel} defined as 
\begin{eqnarray}\label{Eq5}
|\Psi_{12}\rangle=\sqrt{1-\lambda^2}\sum_{n=0}^\infty\lambda^n|n,n\rangle.
\end{eqnarray}
Notice that, here, the number of photons in both channels is given by the parameter $\lambda$, $N=N_2=N_1=\langle \Psi_{12}|a^\dag_1a_1|\Psi_{12}\rangle=\frac{\lambda^2}{1-\lambda^2}$. Then, the signal beam $a_1$ is emitted following the same protocol shown for the coherent state, while the idler beam $a_2$ remains in the lab. Therefore, the transformations of Eqs.~\eqref{Eq1} and~\eqref{Eq2} can be directly implemented, adding the subindex 1 to the signal mode and noting that the idler suffers no transformation $a_2''=a_2$. The main difference here is that, since we are using entangled states, we can implement bipartite measurements in the system that enhance the SNR of the quantum protocol. Specifically, we propose to measure the operator $x_1''x_2''-p_1''p_2''$. The SNR obtained in this quantum protocol can be written as
\begin{widetext}
\begin{eqnarray}\label{Eq6}
\left(\frac{S^2}{\sigma^2}\right)_Q=\frac{4\eta N(N+1)(1-\cos(\phi))^2}{\eta(1+4N(N+1)\cos^2 \phi)+(1-\eta)(2n_{th}N+n_{th}+N+1)}.
\end{eqnarray}
\end{widetext}
As in the previous protocol, the result is equivalent to the introduction of the interaction of the signal beam with the cloaked object before the reflection in the background.

Let us now compare the SNR of the classical and the quantum protocols. Notice that, in each of them, $N$ denotes the average number of photons initially coming from the signal beam. In order to improve the classical protocol, the condition $\left(\frac{S^2}{\sigma^2}\right)_Q>\left(\frac{S^2}{\sigma^2}\right)_C$ should hold,
\begin{figure}[t]
\includegraphics[width=0.7\textwidth]{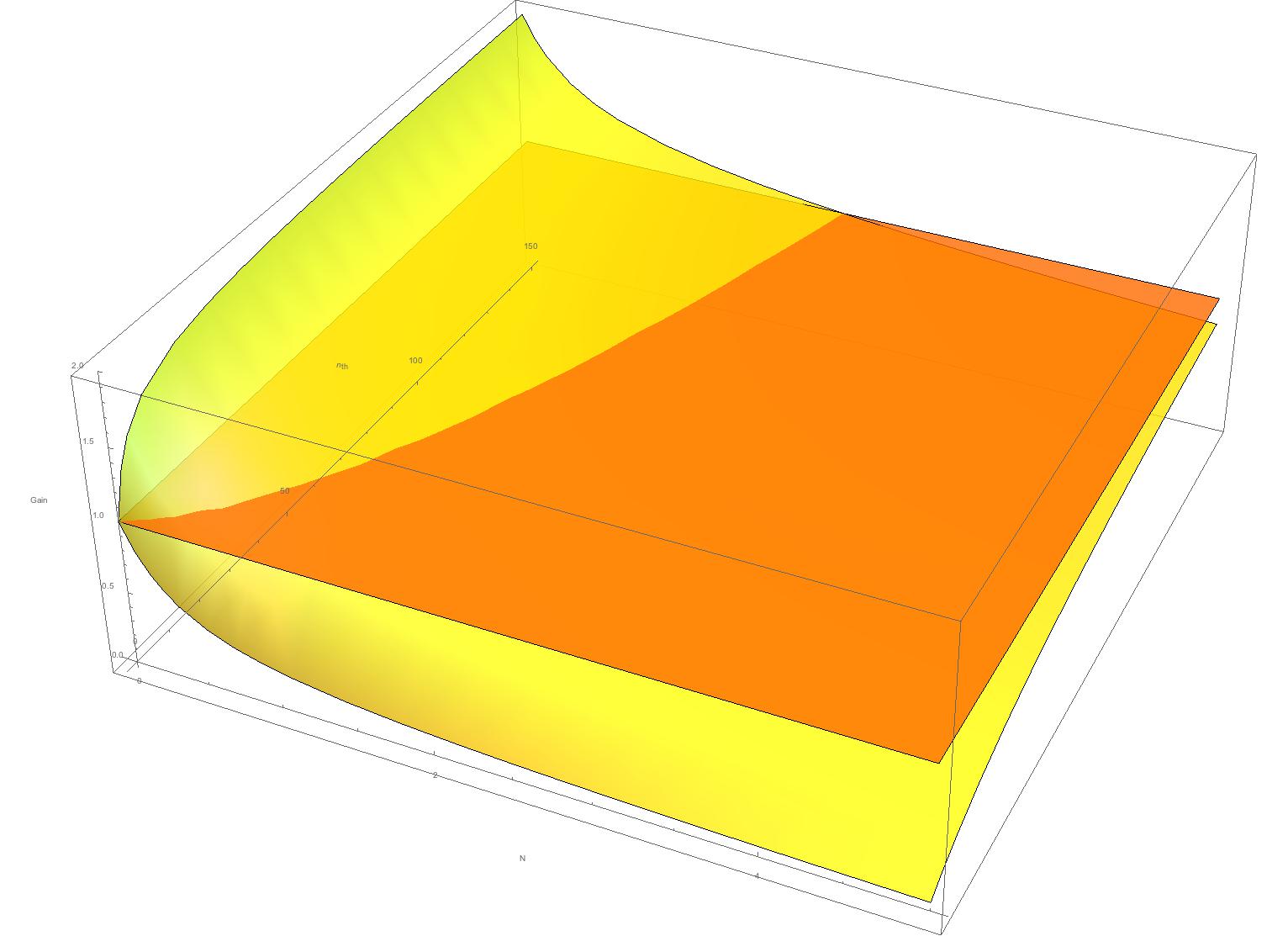}
\caption{This plot represents the inequality given by Eq. \eqref{ineq} in terms of the parameters $N$ and $n_{th}$. The yellow surface represents the left hand side of the inequality for the worst case $\cos \phi = 1$ and a value of the relative reflectivity $\frac{1-\eta}{\eta} = 0.2$. When this surface is above the orange plane, there is a gain of the quantum protocol with respect to the classical one. The larger is the relative reflectivity, the larger is this area.}
\label{GainPlot}
\end{figure}
\begin{eqnarray}\label{ineq}
\frac{\left(\frac{S^2}{\sigma^2}\right)_Q}{\left(\frac{S^2}{\sigma^2}\right)_C}=\frac{(N+1)(1+\frac{1-\eta}{\eta}(2n_{th}+1))}{1+4N(N+1)\cos^2 \phi+\frac{1-\eta}{\eta}(2n_{th}N+n_{th}+N+1)}>1,
\end{eqnarray}
where we have used Eqs.~\eqref{Eq4} and~\eqref{Eq6}. Firstly, let us point out that in the limit $N \gg 1$, there is no improvement (see Fig. \ref{GainPlot}), so we will focus on the limit $N\ll 1$, which corresponds to the case in which we want to detect the target without being detected. Considering the worst scenario $\cos^2\phi=1$, we get the expression
\begin{eqnarray}
4N^2+3N-\frac{1-\eta}{\eta}n_{th}<0.
\end{eqnarray}
Since the equation has to be negative and the parabola has a positive second derivative, the solutions of the inequality are the ones between the roots of the equation with $N>0$,
\begin{eqnarray}
0 < N < \frac{-3+\sqrt{9+16\frac{1-\eta}{\eta}n_{th}}}{8}.
\end{eqnarray}
Therefore, there exists always a finite $N$ enhancing the classical protocol for any phase shift $\phi$, since $\frac{1-\eta}{\eta} n_{th} > 0$. Let us study now the different regimes for a small number of photons $N$ in the two-mode squeezed state. The ratio of the quantum and classical SNR can be approximated as:
\begin{eqnarray} \label{Eq10}
\frac{\left(\frac{S^2}{\sigma^2}\right)_Q}{\left(\frac{S^2}{\sigma^2}\right)_C}&\approx&\frac{1+\frac{1-\eta}{\eta}+2\frac{1-\eta}{\eta}n_{th}}{1+\frac{1-\eta}{\eta}+\frac{1-\eta}{\eta}n_{th}}-\frac{(3+\frac{1-\eta}{\eta}n_{th}) (1+\frac{1-\eta}{\eta}+2\frac{1-\eta}{\eta}n_{th})N}{(1 + \frac{1-\eta}{\eta} + \frac{1-\eta}{\eta}n_{th})^2},
\end{eqnarray}
up to corrections in $O(N^2)$.
Considering the most physical scenario in which $n_{th}\gg1$ is high due to the noisy environment and a highly-reflective background $\frac{1-\eta}{\eta}\sim0$, there are three possible regimes parametrized by the product $\frac{1-\eta}{\eta} n_{th}$. In the first case, in which $\frac{1-\eta}{\eta} n_{th} \gg1$, the gain is $2$, as one may observe from Eq. \eqref{Eq10}. On the other hand, when $\frac{1-\eta}{\eta} n_{th} \ll1$, there is no gain, as discussed above. For intermediate regimes, we may observe that the zero-order term in Eq. \eqref{Eq10} grows monotonously in  $\frac{1-\eta}{\eta} n_{th}$, so one obtains an intermediate gain.

Up to now, we have shown that the measurement in the quantum protocol achieves a gain of up to 2 with respect to the classical protocol. Let us now discuss how to perform this measurement in a microwave technology device. First, we rewrite it in terms of the annihilation operators: $\langle x_1''x_2''-p_1''p_2''\rangle=\langle a_1a_2+a_1^\dag a_2^\dag\rangle$. In order to implement this interaction with photocounters, a Josephson mixer (JM) is required~\cite{Flurin}. The transformation implemented by the JM in the signal and idler modes can then be expressed as
\begin{eqnarray}
a'''_1=\sqrt{G}a''_1+\sqrt{G-1}a_2^{\prime\prime\dag},\label{Eq11}\\
a'''_2=\sqrt{G}a''_2+\sqrt{G-1}a_1^{\prime\prime\dag},\label{Eq12}
\end{eqnarray}
where, following the previous notation, $a_1''(a_2'')$ is the bosonic operator field of the incoming signal (idler) field, $a_1'''(a_2''')$ is the outcoming field and $G>1$ is the chosen gain in the JM, see Fig.~\ref{MWProtocol}. Measuring the operator $O\equiv Ga_2^{\prime\prime\prime\dag}a_2'''-(G-1)a_1^{\prime\prime\prime\dag}a_1'''$, in which the number of photons of signal and idler beams are subtracted with weights corresponding to the JM gain, and therefore are implementable via photocounters, we obtain the following expression:
\begin{eqnarray}\label{Eq13}
\langle O\rangle&=&\langle Ga_2^{\prime\prime\prime\dag}a_2'''-(G-1)a_1^{\prime\prime\prime\dag}a_1'''\rangle=\langle(G-1)+(2G-1)a_2^{\prime\prime\dag}a_2'+\sqrt{G(G-1)}(a_1''a_2''+a_1^{\prime\prime\dag}a_2^{\prime\prime\dag})\rangle\nonumber\\
&=&(G-1)+(2G-1)N+2\sqrt{G(G-1)}\eta\sqrt{N(N+1)}\cos(\phi).
\end{eqnarray}
In order to obtain the ratio between the SNR of the classical protocol and the one involving the JM, we follow the same approach as in Eq.~\eqref{Eq10}. It is straightforward to see that the third term corresponds to the measurement presented above, $\langle x_1''x_2''-p_1''p_2''\rangle$, and since the first term is a constant and the second term is small for $N\ll1$, the improvement with respect to the classical protocol is close to the one of $\langle x_1''x_2''-p_1''p_2''\rangle$, see supplementary information for detailed calculations. 

\begin{figure}[h]
\includegraphics[width=0.5\textwidth]{MWProtocol.jpg}
\caption{Scheme of the quantum protocol for detecting the presence of a cloaked object using microwave technology. First, two squeezed states are mixed with a beam splitter obtaining a two-mode squeezed state Eqs.~\eqref{Eq5}. Secondly, the signal mode is mixed with a thermal state with a high-reflectivity mirror that emulates the background. After this, the object may suffer a phase shift $\phi$ due to the cloaking of the target we want to detect. Then, signal and idler beams are mixed with a Josephson mixer, and the number of photons is finally measured in the signal and idler channels. Red dashed lines show the transformations in the fields due to the elements of the protocol as shown in Eqs.~\eqref{Eq1},~\eqref{Eq2},~\eqref{Eq11}, and~\eqref{Eq12}.}
\label{MWProtocol}
\end{figure}

As can be seen, the implementation of this protocol in microwave technology requires photocounters. Up to date, several proposals have been studied to achieve photodetection in the microwave regime~\cite{Romero,Peropadre,Fan,Sathyamoorthy,Kindel}, which can be extended to photocounters. In this work, we have observed that measuring correlations between the signal and idle beams in a single quadrature provides no gain with respect to the classical protocol and hence, more than one quadrature has to be measured at the same time. Even though we expect that this behavior is generic, we do not have a general proof. The photon number operator provides the complexity of measuring more than one quadrature correlation and its implementation might be the easiest way to perform the measurement.

For completeness, let us analyze the case in which the photocounters are not perfect. The gain of the whole protocol could therefore be defined as the product of the theoretically predicted gain multiplied with the photocounter efficiency. The latter is modeled by adding a beam splitter with reflectivity $\chi$ before the photocounter (see supplementary information). Considering that both photocounters have the same efficiency, one finds
\begin{equation}\label{eq14}
\chi>\frac{1}{2}+\left(2\sqrt{\frac{\eta N}{\epsilon}}\right)\frac{1}{n_{th}(1-\eta)} \,, 
\end{equation}
in order to have an enhancement with respect to the classical protocol, with $\epsilon=G-1>\eta N/(n_{th}(1-\eta))^2$. Let us recall that the expression given by Eq.~\eqref{eq14} is valid in the limit in which $n_{th}(1-\eta) \gg 1$, $N\ll 1$, and $1-\eta \ll 1$ (detailed calculations can be found in the supplementary information). Moreover, the enhancement is actually linear with $\chi$. However, if we take into account photon losses during the freespace path and losses in the JM, a slightly higher efficiency would be required.

Another source of errors is the imperfection of light sources. Indeed, while the generation of coherent states can be achieved with great accuracy, the production of the entangled light is more subtle. This issue has been theoretically and experimentally studied in microwaves in Ref.~\cite{Fedorov2} and it was shown that the main error sources are an imperfect initial vacuum, which means that we start in a thermal state before the JPA, and internal imperfections of the JPA. By studying the resilience of the entanglement with respect to these errors, the conclusions are that it affects mainly to the time window available to perform joint measurements in both beams, while the change in the degree of entanglement could be make negligible with respect to the inefficiency in the photodetection.

Summarizing, in this work we have proposed a quantum illumination protocol to detect cloaked objects, and we have specifically studied the implementation in quantum microwave technology. We have analytically studied the regimes in which there is a gain with respect to the optimal classical protocol by calculating the SNR. Indeed, we have found that the phase shift introduced by cloaking can be detected with a gain of up to 3~dB by employing entangled light beams and joint measurements. Moreover, we have proposed its implementation in microwaves making use of a Josephson mixer and two microwave photocounters. This work demosntrates the potential of quantum technologies, and in particular quantum illumination, for studying the fundamental quantum limits in the field of optical and microwave cloaking. From a technological point of view, it also motivates the development of the long-term missing photocounters in the microwave regime.

{\setlength{\parindent}{0pt}

\section*{AUTHOR CONTRIBUTIONS}
U.L.H., as the first author, has been responsible together with M.S. for the development of this work. U.L.H. and M.S., supported by R.D.C., have made the mathematical demonstrations, carried out calculations, and provided examples. The experimental feasibility in superconducting circuits has been checked by K.F. and F.D. Finally, M.S. and E. S. suggested the seminal ideas. All authors have carefully proofread the manuscript. E.S. supervised the project throughout all stages.

\section*{ACKNOWLEDGMENT}
The authors thank Ryan Sweke, A. Marx and S. Pogorzalek for the useful discussions. We acknowledge support from Spanish MINECO/FEDER FIS2015-69983-P, Basque Government IT472-10, UPV/EHU PhD grant, AQuS project, the German Research Foundation through FE 1564/1-1, Elite Network of Bavaria through the program ExQM, and the IMPRS Quantum Science and Technology. 

\section*{ADDITIONAL INFORMATION} 
The authors declare no competing financial interests.


\begin{thebibliography}{21}

\bibitem{Dolin} Dolin, L. S. On a possibility of comparing three-dimensional electromagnetic systems with inhomogeneous filling. {\it Izv. Vyssh. Uchebn. Zaved. Radiofiz.} {\bf 4}, 964 (1961).

\bibitem{Kerker} Kerker, M. Invisible bodies. {\it J. Opt. Soc. Am.} {\bf 65}, 376 (1975) ; {\bf 65}, 1085 (E) (1975). 

\bibitem{Leonhardt} Leonhardt, U. Optical conformal mapping. {\it Science} {\bf 312}, 1777 (2006).

\bibitem{Pendry} Pendry, J. B., Schurig, D. \& Smith, D. R. Controlling Electromagnetic Fields. {\it Science} {\bf 312}, 1780 (2006).

\bibitem{Schurig} Schurig, D., {\it et al}. Metamaterial electromagnetic cloak at microwave frequencies. {\it Science} {\bf 314}, 977 (2006).

\bibitem{Liu} Liu, Y. \& Zhang, X. Metamaterials: a new frontier of science and technology. {\it Chem. Soc. Rev.} {\bf 40}, 2494 (2010).

\bibitem{Lapine} Lapine, M., Shadrivov, I. V. \& Kivshar, Y. S. Colloquium: Nonlinear metamaterials. {\it Rev. Mod. Phys.} {\bf 86}, 1093 (2014).

\bibitem{Alitalo2} Alitalo, P., Bongard, F., Zurcher, J.-F., Mosig, J. \& Tretyakov, S. Experimental verification of broadband cloaking using a volumetric cloak composed of periodically stacked cylindrical transmission-line networks. {\it Appl. Phys. Lett.} {\bf 94}, 014103  (2009).

\bibitem{Alitalo3} Alitalo, P., Luukkonen, O., Mosig, J. R. \& Tretyakov, S. A. Broadband cloaking with volumetric structures composed of two-dimensional transmission-line networks. {\it Microwave Opt. Technol. Lett.} {\bf 51}, 1627  (2009).

\bibitem{Tretyakov} Tretyakov, S., Alitalo, P., Luukkonen, O. \& Simovski, C. Broadband Electromagnetic Cloaking of Long Cylindrical Objects. {\it Phys. Rev. Lett.} {\bf 103}, 103905  (2009).

\bibitem{Alu} Al\`u, A. \& Engheta, N. Plasmonic and Metamaterial Cloaking: Physical Mechanisms and Potentials. {\it J. Opt. A: Pure Appl. Opt.} {\bf 10}, 093002 (2008).

\bibitem{Alitalo} Alitalo, P., Kettunen, H. \& Tretyakov, S. Cloaking a metal object from an electromagnetic pulse: A comparison between various cloaking technique. {\it J. Appl. Phys.} {\bf 107}, 034905 (2010).

\bibitem{Li} Li, J. \& Pendry, J. B. Hiding under the Carpet: A New Strategy for Cloaking. {\it Phys. Rev. Lett.} {\bf 101}, 203901 (2008).

\bibitem{Lai} Lai, Y., Chen, H., Zhang, Z.-Q. \& Chan, C. T. Complementary Media Invisibility Cloak that Cloaks Objects at a Distance Outside the Cloaking Shell. {\it Phys. Rev. Lett.} {\bf 102}, 093901 (2009).

\bibitem{McCall} McCall, M. W., Favaro, A., Kinsler, P. \& Boardman, A. A spacetime cloak, or a history editor. {\it Journal of Optics} {\bf 13}, 2 (2010).

\bibitem{Jiang} Jiang, W. X. \& Cui, T. J. Radar illusion via metamaterials. {\it Phys. Rev. E} {\bf 83}, 026601 (2011).

\bibitem{Nelson} Nelson, P. A. \& Elliott, S. J. {\it Active Control of Sound} (New York: Academic Press, 1992).

\bibitem{Shchelokova} Shchelokova, A. V., {\it et al}. Experimental realization of invisibility cloaking. {\it Physics-Uspekhi} {\bf 58}, 167 (2015).

\bibitem{Liu15} Liu, R. {\it et al}. Broadband Ground-Plane Cloak. {\it Science} {\bf 323}, 366 (2009).

\bibitem{Zhou} Zhou, F. {\it et al}. Hiding a Realistic Object Using a Broadband Terahertz Invisibility Cloak. {\it Sci. Rep.} {\bf 1}, 78 (2011).

\bibitem{Ergin} Ergin, T., Stenger, N., Brenner, P., Pendry, J. B. \& Wegener, M. Three-Dimensional Invisibility Cloak at Optical Wavelengths. {\it Science} {\bf 328}, 337 (2010).

\bibitem{Lloyd} Lloyd, S. Enhanced Sensitivity of Photodetection via Quantum Illumination. {\it Science} {\bf 321}, 1463 (2008).

\bibitem{Zhuang} Zhuang, Q., Zhang, Z. \& Shapiro, J. H. Optimum mixed-state discrimination for noisy entanglement-enhanced sensing. {\it Phys. Rev. Lett.} {\bf 118}, 040801 (2017).

\bibitem{Saphiro} Shapiro, J. H. \& Lloyd, S., Quantum illumination versus coherent-state target detection. {\it New J. Phys.} {\bf 11}, 063045 (2009).

\bibitem{Tan} Tan, S. H. {\it et al}. Quantum Illumination with Gaussian States. {\it Phys. Rev. Lett.} {\bf 101}, 253601 (2008).

\bibitem{Guha} Guha, S. \& Erkmen, B. I. Gaussian-state quantum-illumination receivers for target detection. {\it Phys. Rev. A} {\bf 80}, 052310 (2009).

\bibitem{Zhang} Zhang, S.-L. {\it et al}. Quantum illumination with photon-subtracted continuous-variable entanglement. {\it Phys. Rev. A} {\bf 89}, 062309 (2014).

\bibitem{Lopaeva} Lopaeva, E. D., {\it et al}. Experimental Realization of Quantum Illumination. {\it Phys. Rev. Lett.} {\bf 110}, 153603 (2013).

\bibitem{Zhang13} Zhang, Z., Tengner, M., Zhong, T., Wong, F. N. C. \& Shapiro, J. H. Entanglement Benefit Survives an Entanglement-Breaking Channel. {\it Phys. Rev. Lett.} {\bf 111}, 010501 (2013).

\bibitem{Zhang15} Zhang, Z., Mouradian, S., Wong, F. N. C. \& Shapiro, J. H. Entanglement-Enhanced Sensing in a Lossy and Noisy Environment. {\it Phys. Rev. Lett.} {\bf 114}, 110506 (2015).

\bibitem{Sanders97} Sanders, B. C., Milburn, G. J. \& Zhang, Z. A robust quantum receiver for phase shift keyed signals. {\it J. Mod. Opt.} {\bf 44}, 1309 (1997).

\bibitem{Muller15} M\"uller, C. R. \& Marquardt, C. Optimal quantum measurements for phase-shift estimation in optical interferometry. {\it New J. Phys.} {\bf 17}, 032003 (2015)

\bibitem{Barzanjeh} Barzanjeh, S., {\it et al}. Microwave Quantum Illumination. {\it Phys. Rev. Lett.} {\bf 114}, 080503 (2015).

\bibitem{Sharkov} Sharkov, E. A., {\it Passive Microwave Remote Sensing of the Earth}. Springer (2003).

\bibitem{Lanzagorta} Lanzagorta, M., {\it Quantum Radar}. Morgan \& Claypool (2011).

\bibitem{Sanz}  Sanz, M., Las Heras, U., Garc\'ia-Ripoll, J. J., Solano, E. \& Di Candia, R. Quantum Estimation Methods for Quantum Illumination. {\it Phys. Rev. Lett.} {\bf 118}, 070803 (2017)

\bibitem{Fedorov} Fedorov, K. G., {\it et al}. Displacement of propagating squeezed microwave states. {\it Phys. Rev. Lett.} {\bf 117}, 020502 (2016).

\bibitem{Fedorov2} Fedorov, K. G., {\it et al}. Finite-time quantum correlations of propagating squeezed microwaves. arXiv:1703.05138 (2017).

\bibitem{DiCandia} Di Candia, R., {\it et al}. Quantum teleportation of propagating quantum microwaves. {\it EPJ Quantum Technology} {\bf 2}, 25 (2015).

\bibitem{Menzel} Menzel, E. P. {\it et al}. Path Entanglement of Continuous-Variable Quantum Microwaves. {\it Phys. Rev. Lett.} {\bf 109}, 250502 (2012).

\bibitem{Flurin} Flurin, E., Roch, N., Mallet, F., Devoret, M. H. \& Huard, B. Generating Entangled Microwave Radiation Over Two Transmission Lines. {\it Phys. Rev. Lett.} {\bf 109}, 183901 (2012).

\bibitem{Romero} Romero, G., Garc\'ia-Ripoll, J. J. \& Solano, E., Microwave Photon Detector in Circuit QED. {\it Phys. Rev. Lett.} {\bf 102}, 173602 (2009).

\bibitem{Peropadre} Peropadre, B., {\it et al}. Approaching perfect microwave photodetection in circuit QED. {\it Phys. Rev. A} {\bf 84}, 063834 (2011).

\bibitem{Fan} Fan, B., Johansson, G., Combes, J., Milburn, G. J. \& Stace, T. M. Nonabsorbing high-efficiency counter for itinerant microwave photons. {\it Phys. Rev. B} {\bf 90}, 035132 (2014).

\bibitem{Sathyamoorthy} Sathyamoorthy, S. R., Stace, T. M. \& Johansson, G. Detecting itinerant single microwave photons. {\it Comptes Rendus Physique} {\bf 17}, 756 (2016).

\bibitem{Kindel} Kindel, W. F., Schroer, M.D. \& Lehnert, K. W. Generation and efficient measurement of single photons from fixed-frequency superconducting qubits. {\it Phys. Rev. A} {\bf 93}, 033817 (2016).

\end{thebibliography}
\end{document}